\documentclass[reprint,aip,apl]{revtex4-1}
\usepackage{natbib}
\usepackage{amsmath}
\usepackage{graphicx}
\usepackage{wrapfig}

\begin{document}
\title{Outlook for inverse design in nanophotonics}

\author{Sean Molesky}
\affiliation{Department of Electrical Engineering, Princeton University, Princeton, NJ, 08544}

\author{Zin Lin}
\affiliation{John A. Paulson School of Engineering and Applied Sciences Harvard University, Cambridge, MA, 02138}

\author{Alexander Y. Piggott}
\affiliation{Ginzton Laboratory, Stanford University, Stanford, CA, 94305}

\author{Weiliang Jin}
\affiliation{Department of Electrical Engineering, Princeton University, Princeton, NJ, 08544}

\author{Jelena Vu\u ckovi\'c}
\affiliation{Ginzton Laboratory, Stanford University, Stanford, CA, 94305}

\author{Alejandro W. Rodriguez}
\affiliation{Department of Electrical Engineering, Princeton University, Princeton, NJ, 08544}
\email{arod@princeton.edu} 


\begin{abstract}
  Recent advancements in computational inverse design have begun to
  reshape the landscape of structures and techniques available to
  nanophotonics. Here, we outline a cross section of key developments
  at the intersection of these two fields: moving from a recap of
  foundational results to motivation of emerging applications in
  nonlinear, topological, near-field and on-chip optics.
\end{abstract}
\maketitle

\noindent
The development of devices in nanophotonics has historically relied on
intuition-based approaches, the impetus for which develops from
knowledge of some a priori known physical effect. The specific
features of such devices are then typically calculated and matched to
suitable applications by tuning small sets of characteristic
parameters. This approach has had a long track record of success,
giving rise to a rich and widely exploited library of templates that
includes multilayer thin films\cite{qian2011stable},
Fabry-Perot\cite{szoke1969bistable} and microring
resonators\cite{xu2008silicon}, silicon
waveguides\cite{klohn1978silicon,politi2008silica}, photonic
crystals\cite{Joannop}, plasmonic
nanostructures\cite{anker2008biosensing}, and nanobeam
cavities\cite{eichenfield2009optomechanical}, top of Fig.~1. Combining
the principles of index guiding and bandgap engineering, along with
material resonances, this collection of designs enables remarkable
manipulation of light over bands of frequencies spanning from the
ultra-violet to the mid infrared: group velocity can be slowed by more
than two orders of magnitude\cite{baba2008slow}, light confined to
volumes thousands of times smaller than its free-space
wavelength\cite{caldwell2014sub}, and resonances made to persist in
micron sized areas for tens of millions of
cycles\cite{spillane2005ultrahigh}.\\ \\
Yet, as the scope of nanophotonics broadens to include large bandwidth
or multi-frequency applications, nonlinear phenomena, and dense
integration, continuing with this prototypical approach poses a
challenge of increasing complexity. For instance, consider the design
of a wavelength-scale structure for enhancing nonlinear
interactions\cite{hashemi2009nonlinear}, discussed below. Even in the
simplest case, several interdependent characteristics must be
simultaneously optimized, among which are large quality factors at
each individual wavelength and nonlinear overlaps, which must be
controlled in as small a volume as possible. In such a situation, the
templates of the aforementioned standard library offer no clear or
best way to proceed; there is no definite reason to expect that an
optimal design can be found in any of the traditional templates, or
that such a design necessarily exists. Moreover, the performance of a
given nonlinear device is likely to be highly dependent on the
particular characteristics of the problem, and as greater demands are
placed on functionality it becomes increasingly doubtful that any one
class of structures will have the broad applicability of past devices.
This lack of evident strategies for broadband applications also brings
to attention the space of structures included in the standard photonic
library. Predominately, traditional designs are repetitive mixtures
and combinations of highly symmetric shapes described by a small
collection of parameters. Since intuition-based optimization is then
carried out in terms of these parameters without a great deal of
additional computational effort, bearing aside known bounds based on
fundamental principles\cite{Yu10,miller2015shape,arabi17} typically
little is known about how close any one particular device comes to
performance limits or how it compares to modified design
possibilities.\\ \\


\begin{figure*}[ht!]
  \centering
  \includegraphics{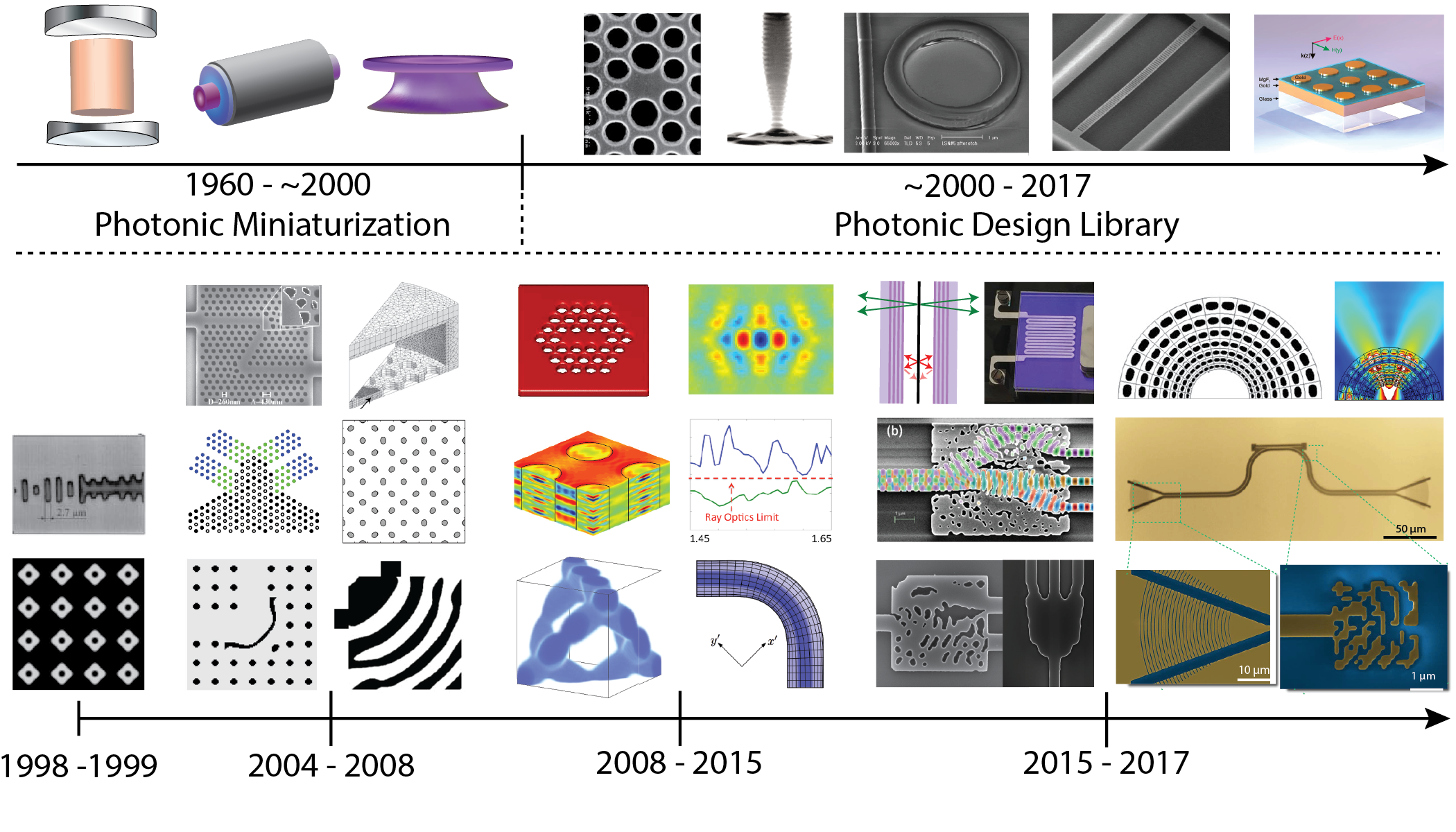}
  \vspace{0 pt}
  \caption{\textbf{Progression of photonic design templates}: (Top)
    During the second half of the twentieth century, advancement in
    fabrication capabilities allowed photonic engineering to expand
    into the micro and nanoscale. Over the past two decades, this
    capability has led to the growth of a rich standard library of
    photonic designs. Moving from left to right, the examples shown
    for photonic miniaturization depict a Fabry-Perot
    cavity\cite{szoke1969bistable}, microdisk
    resonator\cite{levi1993directional}, and fiber
    cavity\cite{ball1992continuously}. The examples for the photonic
    library are a photonic crystal defect cavity from Painter et
    al.\cite{painter1999two}, a micropost cavity from Pelton et
    al.\cite{pelton2002efficient}, a microring resonator from Xu et
    al.\cite{xu2008silicon}, a nanobeam from Eichenfield et
    al.\cite{eichenfield2009optomechanical}, and a plasmonic sensor
    from Liu et al.\cite{liu2010infrared}. (Bottom) The lower part of
    the figure provides a visual companion to the timeline of
    developments in photonic optimization described in the
    text. Working from top to bottom, left to right, the images are
    taken from: (1998-1999) Sp{\"u}hler et al.\cite{spuhler1998very},
    and Dobson and Cox\cite{dobson1999maximizing}; (2004-2008) Borel
    et al.\cite{borel2004topology}, H{\aa}kansson and
    S{\'a}nchez-Dehesa\cite{haakansson2005inverse}, Jensen and
    Sigmund\cite{jensen2004systematic}, Frei et
    al.\cite{frei2008optimization}, Kao et al.\cite{kao2005maximizing}
    and Tsuji and Hirayama\cite{tsuji2008design}; (2008-2014) Lu et
    al.\cite{lu2011inverse} (row), Alaeian et
    al.\cite{alaeian2012optimized}, Men et al.\cite{men2014robust},
    Ganapati et al.\cite{ganapati2014light}, and Liu et
    al.\cite{liu2013transformation}; (2015-2017) Ilic et
    al.\cite{ilic2016tailoring} (row), Fresselen et
    al.\cite{frellsen2016topology}, Piggot et
    al.\cite{piggott2015inverse,piggott2017fabrication} (combined),
    Otomori et al.\cite{otomori2017topology} (row), and Yu et
    al.\cite{yu2017genetically}.}
  \vspace{0 pt}
\end{figure*}

The ability to produce and evaluate novel devices platforms based on
nonlinear and broadband processes, such as solar energy
conversion\cite{alaeian2012optimized,ganapati2014light,xiao2016diffractive},
thermal energy manipulation\cite{ilic2016tailoring,jin2017overcoming},
and on-chip
integration\cite{liu2013transformation,frellsen2016topology,su2017inverse},
will objectively impact the future of nanophotonics.  If the total
design space performance of a given type of design can be even
partially characterized, an immense amount of research effort can be
saved, and a new approach for investigating fundamental limits of
nanophotonic devices could emerge.  In this review, we bring attention
to a collection of recent results showcasing the usefulness of
computational inverse design approaches for both comparing the
relative performance of possible structures, and creating photonic
devices in instances where traditional intuition-based strategies
prove difficult to implement.  We begin by providing background on
inverse design in nanophotonics, highlighting some of the major
developments in this field. From this basis of understanding we then
turn to discussion of emerging applications and experimental
challenges, motivating ways in which inverse design techniques have
and could be employed in nonlinear, topological, near-field, and
integrated optics.

\section{Background} 
\label{sec:background}
\noindent
\textbf{1998}--\textbf{2003}: The driving motivations behind inverse
design have been present for at least several hundreds of years. They
are part of the same family of ideas that led Bernoulli to consider
the brachistochrone problem, Maupertuis to propose the principle of
least action, and Ambartsumian to question the relation between a set
of eigenvalues and its generating differential
equation\cite{chadan2012inverse}. There are at least two central
thrusts: first, to determine the extent that the characteristics of a
solution, either actual or desired, determine the system from which
they are derived; and second, to find effective algorithms for working
from solutions characteristics to physical systems.\\ \\
In the context of nanophotonics, inverse-problem formulations are
understandably much more recent\cite{bendsoetopology}\footnote{For an
  overview of topology optimization in the field of mechanics, where
  many of the techniques now used in nanophotonics were developed, see
  Bends{\o}e and Sigmund\cite{bendsoetopology}. We also note that the
  material presented here is in no way an exhaustive or definitive
  history of nanophotonic optimization. Rather, it is meant to serve
  as a cross section of results giving a sense of how the field has
  evolved. In particular, we will not discuss the closely related
  development of sensitivity analysis\cite{georgieva2002feasible} in
  the microwave community, even though conceptually there is almost no
  difference between these two areas. We direct readers interested in
  a more thorough historical accounts to the reviews by
  Jensen\cite{jensen2011topology} and
  Sigmund\cite{sigmund2011usefulness}, and the articles contained
  therein. Earlier articles by Boa and Friedman or Dobson could also
  be considered as a starting points for gradient based inverse design
  in nanophotonics. Similarly, genetic algorithms had already been
  applied to several problems in electromagnetics, which could be
  considered near enough to nanophotonics. However, within the field
  the two works cited in the main text have had the largest impact.}.
This offshoot, on which we will focus exclusively, began in the late
90s with the work of Sp\"{u}hler et al.\cite{spuhler1998very} and Cox
and Dobson\cite{dobson1999maximizing}, beginning of Fig.~1. In the
first article, Sp\"{u}hler et al. designed and fabricated a
SiO$_{2}$/SiON telecom-fiber to ridge-waveguide coupler. Using a
genetic algorithm to determine the width of the SiON core over a
distance of $138 \mu$m in $3 \mu$m steps, a 2 dB increase in
efficiency was achieved compared to direct coupling. In the second
article, Cox and Dobson applied a gradient-search algorithm to the
problem of bandgap optimization: starting from a 2d periodic structure
composed of two materials, they sought to enlarge its bandgap by
symmetric alterations of the dielectric composition, demonstrating a
34\% increase. The methods used to perform structural optimization in
these two early applications of photonic inverse design stand as
archetypes for classification, involving either genetic
(evolutionary)\cite{back1997evolutionary} or
gradient-based\cite{fu2005simulation} approaches. Crucially, in
genetic algorithms the sensitivity of the fitness or design objective
to the individual design parameters (derivative information of the
objective function) is not necessarily determined. Moreover, even if
gradient information is incorporated into any of the subroutines, it
does not deterministically drive the algorithm. This alteration offers
both benefits and drawbacks. For complex, non-convex objectives the
algorithm is less likely to spend many iterations in oscillatory
regions of the parameter space lacking strong maxima. In exchange, it
is more likely, depending on the problem, that locally optimal designs
are missed and that additional iterations will be required to achieve
convergence comparable to a gradient based approach.\\ \\
\noindent
\textbf{2004}--\textbf{2008}: In the five-year period following these
initial investigations, notable extensions and contributions were
made. Among them, Doosje et al. showed that plane wave expansions
could be used to implement inverse calculations of 3d fcc photonic
crystals\cite{doosje2000photonic}; Cox and Dobson successfully
extended their original work to include in-plane electric
fields\cite{cox2000band}; Felici and Heinz considered optimal designs
for coupling fibers to adiabatic tapers, making use of cascading
algorithms combining coarse- and fine-grained
parameterizations\cite{felici2001shape}; Geremia et
al.\cite{geremia2002inverse} formulated the design of photonic-crystal
cavities as a Lagrangian maximization problem involving a generalized
cost functional defined in terms of the desired optical
characteristics; Jiang et al.\cite{jiang2003parallel} used a genetic
algorithm to achieve mode matching between photonic-crystal and fiber
waveguides; Kizilats\cite{kiziltas2003topology} et al. applied
optimization techniques to improve the design of radio frequency patch
antennas. With few exceptions, most works focused on two classes of
problems, involving either bandgap optimization in photonic crystals
or mode coupling in waveguide
geometries\cite{erni2000application,felici2001shape}. A commonality
among these problems was a high degree of symmetry and low
dimensionality associated with the optimization parameters, with
gradient search methods mainly applied to periodic structures
optimized over a small selection of parameters within a predetermined
family of designs. Large-scale optimization methods nevertheless begun
to be simultaneously pursued in the closely related area of
sensitivity analysis\cite{veronis2004method,jiao2006systematic},
illustrating huge speed-ups in the characterization of the impact of
defects and roughness on photonic devices.\\ \\



The works of Jensen, Sigmund et
al.\cite{borel2004topology,jensen2004systematic,jensen2005topology};
Bruger, Kao, Osher, and
Yablonovitch\cite{burger2003framework,burger2004inverse,kao2005maximizing};
H{\aa}kansson, S{\'a}nchez-Dehesa, and Sanchis et
al.\cite{sanchis2004integrated,haakansson2005inverse}; and Preble,
Lipson\cite{preble2005two} between 2004 and 2005 began to clearly
reshape this landscape, second grouping of Fig.~1. First, inverse
methods were extended to include a wider range of technologically
relevant applications, including photonic-crystal waveguide bends
showing sub 1\% transmission losses over a broad band of
frequencies\cite{jensen2004systematic}, few wavelength-thick devices
capable of acting as frequency
demultiplexers\cite{haakansson2005inverse}, and more varied photonic
crystal configurations for the creation of wide
bandgaps\cite{preble2005two,kao2005maximizing}. Second, the
introduction of adjoint
topology\cite{borel2004topology,jensen2004systematic} and level-set
optimization along with improvements to existing genetic techniques,
vastly broadened the generality and computational efficiency of
inverse design.\\ \\
At a high level, a major benefit of introducing the concepts of
level-set and topology optimization is that they provide a systematic
way to organize design possibilities. In the level-set method, a given
design is described by partitioning the physical optimization domain
$D$ into level sets of a solution function $\Phi(\mathbf{x})$ that
varies continuously over space $\mathbf{x} \in D$ (defined over each
voxel or mesh in a computational cell). Mimicking the description of
Bruger et al.\cite{burger2004inverse}, to consider smooth candidate
structures consisting of two materials, one can define a partitioning,
\begin{align}
  \Omega_{1} &= \left\{\Phi\left( \mathbf{x}\right) < \text{low}
  \right\}, \nonumber \\ \Omega_{2} &= \left\{\text{high} < \Phi\left(
  \mathbf{x}\right) \right\},
\end{align} 
that maps an otherwise continuously varying function to a binary
domain. To move toward a device design, $\Phi\left(\mathbf{x}\right)$
is then evolved either through an equation of motion (such as the
Hamilton-Jacobi equation) or via gradients\cite{van2013level}, causing
it to settle at local maxima. Specifying the optimization domain in
this way allows for floating boundaries between material components
without the need to provide explicit parameterizations. Additionally,
it also allows for appearance of voids while mitigating conditions
conducive to the development of ultra-fine (pixel checkerboard)
features\cite{van2013level}.
\\ \\
In topology optimization\footnote{Note that technically speaking, the
  level set method along with various other shape optimization
  approaches, is a subset of topology optimization. Here, by topology
  optimization we mean the class of voxel-based optimization
  algorithms typically referenced in nanophotonics.} an even broader
design space is considered. Drawing from the finite discretization of
the underlying physical problem in a numerical method, each node (line
segment, pixel or voxel) within a computational cell is treated as a
degree of freedom and ``relaxed'' continuously in some
range. Mimicking the implementation of Jensen and
Sigmund\cite{jensen2005topology} as an example, the permittivity of
each node $\epsilon_i$ in structures consisting of two materials is
defined as a linear functional,
\begin{equation}
   \epsilon_i = \epsilon_{1} + \lambda_{i}\left(\epsilon_{2} -
   \epsilon_{1}\right),
   \label{epsTopoOpt}
\end{equation}
where $\epsilon_{1,2}$ denotes the permittivity of the two materials
and $\lambda \in \left[ 0, 1\right]$ acts as a relaxation
parameter\footnote{The example given here is only one of many such
  formulations that have been considered\cite{tsuji2008design}.}. The
problem of finding an optimal structure over the space of all
discretized designs then amounts to determining the value of $\lambda$
for each node, while ensuring that the latter takes only its extreme
values.\\ \\
In either approach, the space of possible designs is enormous (on the
order of the set of nodes in the optimization domain) and hence to
ensure any hope of convergence to a local optimum, iterations based on
the gradient of the objective with respect to the design parameters
are needed\cite{sigmund2011usefulness}. Here, gradients provide the
optimization algorithms a direction of improvement; remarkably, while
there is no provable guarantee of a globally optimal solution, it is
still nevertheless possible to find good (and in some cases globally
optimal\cite{Miller14}) designs. The computational effort required to
determine and make use of this information is made manageable through
the use of the adjoint
method\cite{mbgiles_ftc2000,lalau-keraly_oe2013}, described in Box 1.\\ \\
\noindent 
\onecolumngrid \phantom \\
\noindent
\textbf{Box 1: The adjoint method} \\
\noindent
\rule{1.0\textwidth}{1.0 pt}
\twocolumngrid 

\begin{figure}[t]
  \centering
  \vspace*{0.3cm}\includegraphics{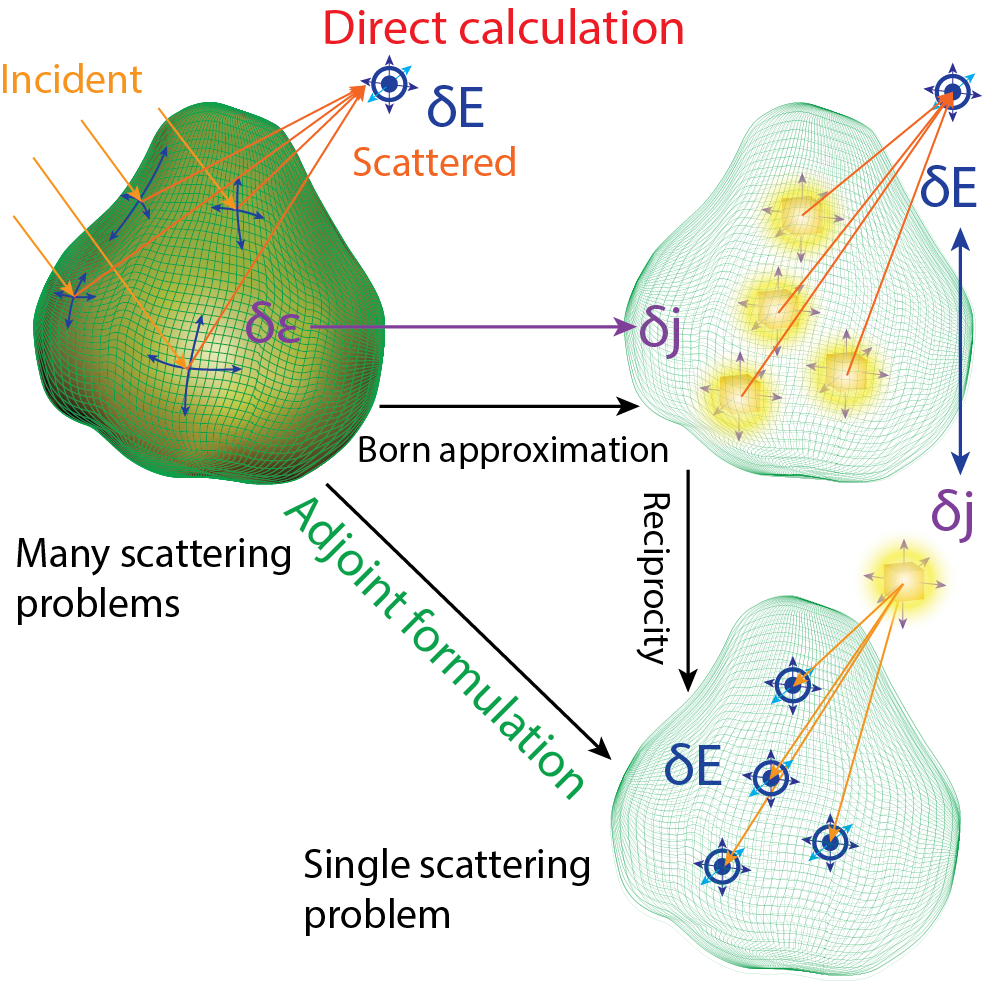}
\end{figure}

Let
$\mathcal{F}\left[\psi\left(\textbf{x}\right),\epsilon\left(\textbf{x}\right)\right]$
be some objective functional, $\psi\left(\textbf{x}\right)$ a field
that $\mathcal{F}$ is optimized relative to,
$\epsilon\left(\textbf{x}\right)$ a controllable set of design
parameters, and
$\overline{\mathcal{M}}\left[\psi\left(\textbf{x}\right),\epsilon\left(\textbf{x}\right)\right]=0$
a collection of constraints between $\psi\left(\textbf{x}\right)$ and
$\epsilon\left(\textbf{x}\right)$, with $\textbf{x}$ parameterizing
the computational domain. The relevant derivative (sensitivity)
information for locally optimizing $\epsilon\left(\textbf{x}\right)$
is then given by the total variation of $\mathcal{F}$ with respect to
the design parameters:
\begin{equation}
\delta_{\epsilon\left(\textbf{x}\right)}\mathcal{F}=\frac{\delta\mathcal{F}}{\delta \epsilon\left(\textbf{x}\right)}+\frac{\delta\mathcal{F}}{\delta \psi\left(\textbf{x}\right)}\frac{\delta \psi\left(\textbf{x}\right)}{\delta \epsilon\left(\textbf{x}\right)}. 
\label{optimizationEq} 
\end{equation}
Since there is only one objective functional, $\delta\mathcal{F} /
\delta \epsilon\left(\textbf{x}\right)$ and $\delta\mathcal{F} /
\delta \psi\left(\textbf{x}\right)$ can be dealt with in a
straightforward way. The determination of $\delta
\psi\left(\textbf{x}\right)/\delta \epsilon\left(\textbf{x}\right)$ is
more difficult. Functionally, this quantity is defined by the equation,
\begin{equation}
  \delta_{\epsilon\left(\textbf{x}\right)}\overline{\mathcal{M}}=\frac{\delta\overline{\mathcal{M}}}{\delta \epsilon\left(\textbf{x}\right)}+\frac{\delta\overline{\mathcal{M}}}{\delta \psi\left(\textbf{x}\right)}\frac{\delta \psi\left(\textbf{x}\right)}{\delta \epsilon\left(\textbf{x}\right)}=0,
  \label{constraintEq}  
\end{equation}
giving $\delta \psi\left(\textbf{x}\right)/\delta
\epsilon\left(\textbf{x}\right)=
-\left(\delta\overline{\mathcal{M}}/\delta
\psi\left(\textbf{x}\right)\right)^{-1}\delta\overline{\mathcal{M}}/\delta
\epsilon\left(\textbf{x}\right)$. This expression is computationally
costly, since $\mathcal{M}$ is typically determined by the solution of
the underlying physical problem, or the inverse of the system
matrix. Treating it in the forward direction requires as many
solutions as there are optimization unknowns. Treating it in the
reverse requires first computing the matrix inverse and then
applying/using the resulting dense matrix.\\ \\
To avoid such a calculation, which would severely limit the
type of problems that could be tractably considered, one typically
substitutes the constraint vector for the objective
$\mathcal{F}$. Inserting \eqref{constraintEq} into
\eqref{optimizationEq} one obtains,
\begin{equation}
  \delta_{\epsilon\left(\textbf{x}\right)}\mathcal{F}=\frac{\delta\mathcal{F}}{\delta \epsilon\left(\textbf{x}\right)}-\frac{\delta\mathcal{F}}{\delta \psi\left(\textbf{x}\right)}\left(\frac{\delta\overline{\mathcal{M}}}{\delta \psi\left(\textbf{x}\right)}\right)^{-1}\frac{\delta\overline{\mathcal{M}}}{\delta \epsilon\left(\textbf{x}\right)},
  \label{interMediate}
\end{equation}
in which case the combination
$\overline{\lambda}_{\textbf{x}}\left(\cdot\right)=\delta\mathcal{F} /
\delta\psi\left(\textbf{x}\right)\left(\delta\overline{\mathcal{M}} /
\delta \psi\left(\textbf{x}\right)\right)^{-1}$ acts as a linear
functional on $\delta\overline{\mathcal{M}} /\delta
\epsilon\left(\textbf{x}\right)$. This gives the adjoint
equation\footnote{This result can also be deduced from the method of
  Lagrange multipliers. Setting
  $\mathcal{H}\left(\psi\left(\textbf{x}\right),\epsilon\left(\textbf{x}\right)\right)
  =
  \mathcal{F}\left(\psi\left(\textbf{x}\right),\epsilon\left(\textbf{x}\right)\right)
  -\overline{\lambda}_{\textbf{x}}\left(
  \overline{\mathcal{M}}\left(\psi\left(\textbf{x}\right),\epsilon\left(\textbf{x}\right)\right)\right)$
  \eqref{adjointMethod} is found to result from
  $\delta_{\psi\left(\textbf{x}\right)}\mathcal{H}$, and the
  $\delta_{\epsilon\left(\textbf{x}\right)}$ variation of the
  constraint equation.}
\begin{equation}
  \left(\frac{\delta\overline{\mathcal{M}}}{\delta \psi\left(\textbf{x}\right)}\right)^{\dagger}\overline{\lambda}\left(\textbf{x}\right)=\frac{\delta\mathcal{F}}{\delta \psi\left(\textbf{x}\right)},
  \label{adjointMethod}
\end{equation}
\noindent
\textbf{2008}--\textbf{2015}: Following the first forays of
large-scale optimization methods in photonics was a contemporaneous
push to investigate structures and applications of increasing
complexity, including early works in solar energy
harvesting\cite{alaeian2012optimized,ganapati2014light}, dispersion
engineering\cite{riishede2008inverse}, wavelength
focusing\cite{dobson2009optimization}, and nonlinear
switching\cite{elesin2012design}. The corresponding gains in
performance determined by adjoint techniques naturally led to
questions concerning the incorporation of realistic constraints and
computationally workable extensions to larger design domains, third
grouping of Fig.~1. In essence, while increasing generality provided a
boon to device performance, these gains nevertheless came at a
cost. \noindent In the absence of constraints, the feature sizes that can be
produced by either level-set or topology optimization methods are
limited only by the size of the chosen computational
domains. Moreover, in the topology optimization approach, the
permittivity is allowed to vary continuously to make use of gradients
and depending on how material constraints are implemented, the solution of which yields all the required derivative
information. Here, $\dagger$ is the adjoint operator. If the
constraint equations are linear in $\psi\left(\textbf{x}\right)$, then
$\delta\overline{\mathcal{M}}/\delta \psi\left(\textbf{x}\right)$ is
just the operator form of the constraint vector, and the same
factorization and conditioners used to solve for
$\psi\left(\mathbf{x}\right)$ can be used to reduce the cost of
determining $\lambda\left(\textbf{x}\right)$.\\ \\
It is remarkable that adjoint methods yield full derivative
information from the solution of a problem that is, in every respect,
no harder to solve than the original problem. This fact can, in
certain situations, be understood intuitively. Consider for instance
the typical electromagnetic problem of optimizing scattered power from
a body due to some known incident field. Such a problem requires
determining the sensitivity or change in scattered power due to
changes in the permittivity $\delta \epsilon$ throughout the body,
illustrated in the accompanying schematic. The direct or ``forward''
approach to the problem proposed by \eqref{constraintEq} goes as
follows: First, one can exploit the Born approximation to treat the
first-order variation in the polarization $\delta j = \delta \epsilon
\times (\text{incident field})$ at each position as an independent
(induced) current source. To determine the resulting change in power
due to $\delta \epsilon$, a field calculation needs to be carried out
at every position in the body (requiring as many solutions as there
are polarization unknowns). Then, one must take the inner product
between each resulting field and the initial scattered field. However,
electromagnetic reciprocity\cite{Joannop} implies that the roles of
the field and source can be interchanged, which right away yields the
``backward'' problem. Applying this transformation to the initial
scattered field (no variations), the sensitivity calculation is recast
as the problem of determining the field inside the object caused by a
single source outside the body, requiring the solution of a single
``adjoint'' scattering problem, described by
\eqref{adjointMethod}. Remarkably, while this concrete example
provides intuition, the power of the adjoint method is its generality,
applying to a much broader set of problems, e.g. nonreciprocal and
even nonlinear materials\cite{mbgiles_ftc2000}.  \onecolumngrid
\noindent
\phantom \\ \\
\rule{1.0\textwidth}{1.0 pt}
\phantom \\ \\
\twocolumngrid
\noindent this can result in intermediate ``gray'' structures, in
which case many iterations are spent considering designs with
graded-index features before settling on binary or piecewise-constant
structures. Dealing with such issues amounts to finding the correct
point of trade-off in searching the design space.  Advancements taking
place between 2008 and 2015 provided foundational understanding of how
such a trade-off occurs and how the imposition of penalization filters
on either objectives or degrees of freedoms can impact device
performance.

\begin{figure*}[t!]
  \centering
  \includegraphics{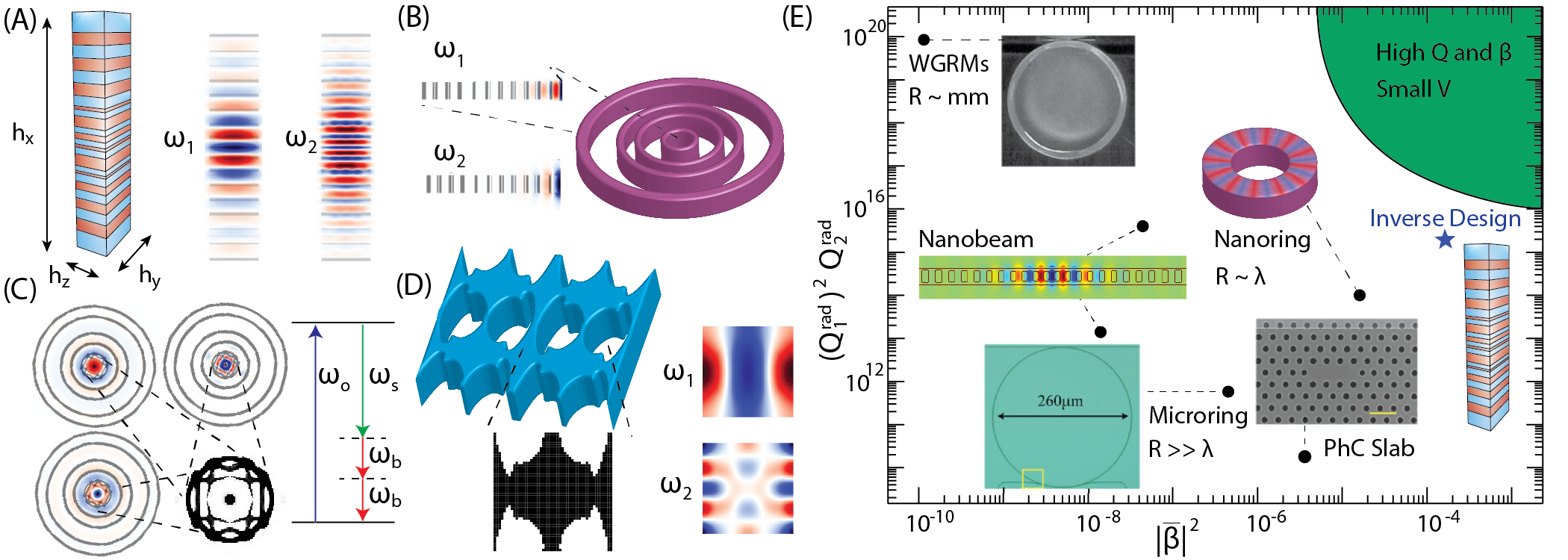}
  \vspace{0 pt}
  \caption{\textbf{Nonlinear optics}: Nonlinear optical interactions
    in micro and nanoscale resonators are regulated by the modal
    quality factors and nonlinear overlap of the participating
    modes. Even in the simplest of processes, envisioning structures
    that optimally control and select from this parameter space is
    challenging. Shown in the figure are three initial applications of
    inverse design towards this problem. (A) Schematic of a micropost
    cavity consisting of aperiodically alternating AlGaAs/AlO$_{X}$
    layers designed to enhance the efficiency of $\chi^{2}$ second
    harmonic generation\cite{lin2016cavity}. (The figures of merit of
    this design are depicted and compared in (E).) (B) and (C)
    Topology optimized gallium arsenide multi-track ring
    resonators\cite{lin2017topology} clad in silica for $\chi^{2}$
    second harmonic generation, (B), and $\chi^{3}$ difference
    frequency generation, (C). (D) A gallium phosphide metasurface
    designed for $\chi^{2}$ second harmonic
    generation\cite{sitawarin2017inverse}. All designs are found to
    have respective nonlinear figures of merit between one and three
    orders better than previously reported designs.}
  \vspace{0 pt}
\end{figure*}
In 2008, Tsui and Hirayama\cite{tsuji2008design} investigated the
particular problem of losses at a $90^{o}$ bend, showing that the
replacement of the linear relaxation parameter in \eqref{epsTopoOpt}
with a smooth function approaching a step discontinuity yields similar
convergence properties and structures as those obtained using
established penalization methods\cite{sigmund1998numerical}. In a
similar spirit, Wang et al.\cite{wang2012high} studied basic tradeoffs
associated with applications of topology and few-parameter
optimization methods to the realization of slow-light photonic-crystal
waveguides, showing that while the performance of structures generated
by topology optimization is typically superior, producing group
velocity indices approaching 300, similar order of magnitude gains can
be achieved via simple shape variations.  Simultaneously, adaptations
of large-scale methods to include fabrication tolerances were pursued
by Sigmund\cite{sigmund2009manufacturing} generally and by Oskooi et
al.\cite{oskooi2012robust} in the context of robust waveguide tapers,
with the key goal being to produce designs that provide some degree of
optimality while remaining robust with respect to structural
variations. In the former, this was done by introducing erosion and
dilation operators that alter the optimized structure; in the latter,
this was done by considering the performance of the structure relative
to alterations in the direction of steepest descent of the objective
function. To control feature sizes within the level-set method, a
typical approach is to exploit shape parameterizations that
automatically satisfy the desired minimum feature constraint, known as
a geometry projection method.  For instance, Frei, Johnson et
al.\cite{frei2007geometry} observed that a truncated expansion of the
level-set function in a basis of radial functions implicitly
determines the smallest feature size in a given problem, which they
applied to demonstrate a tripling of the Purcell factor of a
single-defect photonic crystal cavity. Practical evaluations of the
relative importance of different facets of topology optimization and
level-set algorithms also extend to computational efficacy, and key
ideas emerged in the works of Lu et al., Men et al., Liang and
Johnson, and Liu et al.\cite{liu2013transformation}, summarized below.\\ \\
\noindent
\textbf{Relaxation methods}: In general, the inverse problem of
determining both field and structural unknowns can be broken into two
subproblems, explored in Lu et al\cite{lu2011inverse}. First, one
relaxes the absolute constraints imposed by Maxwell's equations and
determines the electric field that simultaneously minimizes both the
objective and residual with respect to solution of Maxwell's
equations. Next, one considers the electric field found in the first
step as a given and seeks instead to minimize said residual by solving
for the correct permittivity\cite{lu2011inverse}. An optimal structure
is produced by alternating between these two, less demanding
subproblems, known as a relaxation method. Similar notions were used
earlier by Geremia et al.\cite{geremia2002inverse} and Englund et
al.\cite{Englund05} in inverse design studies of defect cavities.\\ \\
\noindent
\textbf{Subspace methods:} In cases where it is feasible to determine
the modes most actively dictating the objective function, then the
total optimization problem can be limited to more tractable
subspaces\cite{men2010bandgap}. If the allowed variations between any
two iterations is also small, then exploiting a relaxation method as
above allows the modes of one structure to be used as an approximation
for the modes of another; combining these two ideas results in a
considerably more accommodating system of equations that can be solved
by semi-definite programming techniques, explored by Men et
al.\cite{men2010bandgap} in the context of bandgap optimization. Along
a similar vein, the computational cost of problems involving multiple
frequency bands can be dramatically reduced by the use of window
functions and complex-frequency deformations, so long as the objective
function is analytic\cite{liang2013formulation}. First, multiply the
objective by a meromorphic function peaked around the frequency bands
of interest (a Lorentzian is given as an example). By analytically
continuing to the complex plane, the entire integral objective is
obtained from the residues of the window function, requiring fewer
calculations. In Men et al., this formulation was applied to produce
3d cavities with Purcell factors larger than $10^{5}$.\\ \\
\noindent
\textbf{Transformation optics}: Finally, the strengths of inverse
design and transformation optics are highly
complementary\cite{liu2013transformation}. Coordinate transformations
often lead to a clear understanding of the boundary behavior that must
be achieved for a device to perform efficiently. For example, in order
to keep modes from scattering around a waveguide bend, any
permittivity profile that has the same effect as a coordinate
transformation producing a $90^{o}$ rotation will work. However,
natural transformations, such as a change to circular coordinates in
the above example, tend to produce material profiles that are
difficult to fabricate, (unrealistic permittivities, anisotropy,
etc). On the other hand, this second problem is well suited for
inverse design. Knowledge of the exact boundary conditions that must
be satisfied means that Maxwell's equations do not need to be solved
at each iteration. Instead, the algorithm may focus solely on the
objective function, like minimization of anisotropy in the
permittivity, requiring only derivative information.\\ \\
Results stemming from these computational insights are also
intriguing. For instance, Men et al.\cite{men2014robust} found that
even without imposing fabrication constraint, their inverse design
algorithm could not find photonic structures with fractional bandgaps
larger than $\approx\; 30\%$ (for index contrasts smaller than
1:3.6). Perhaps surprisingly, such gaps are only slightly larger than
those previously attributed to hand-designed fcc photonic
crystals\cite{maldovan2004diamond}.  Considering the large number of
degrees of freedom and initial designs explored, the results suggest
that there is not much room for further bandgap engineering. The
suggestion of such a fundamental limitation, while negative, is quite
appealing from a theoretical perspective. Instinctively, the size of
bandgaps must be in some way ultimately bound by material constraints,
regardless of how these materials are spatially distributed. Yet, the
existence of an argument proving this fact remains open.

\begin{figure}
  \centering
  \includegraphics{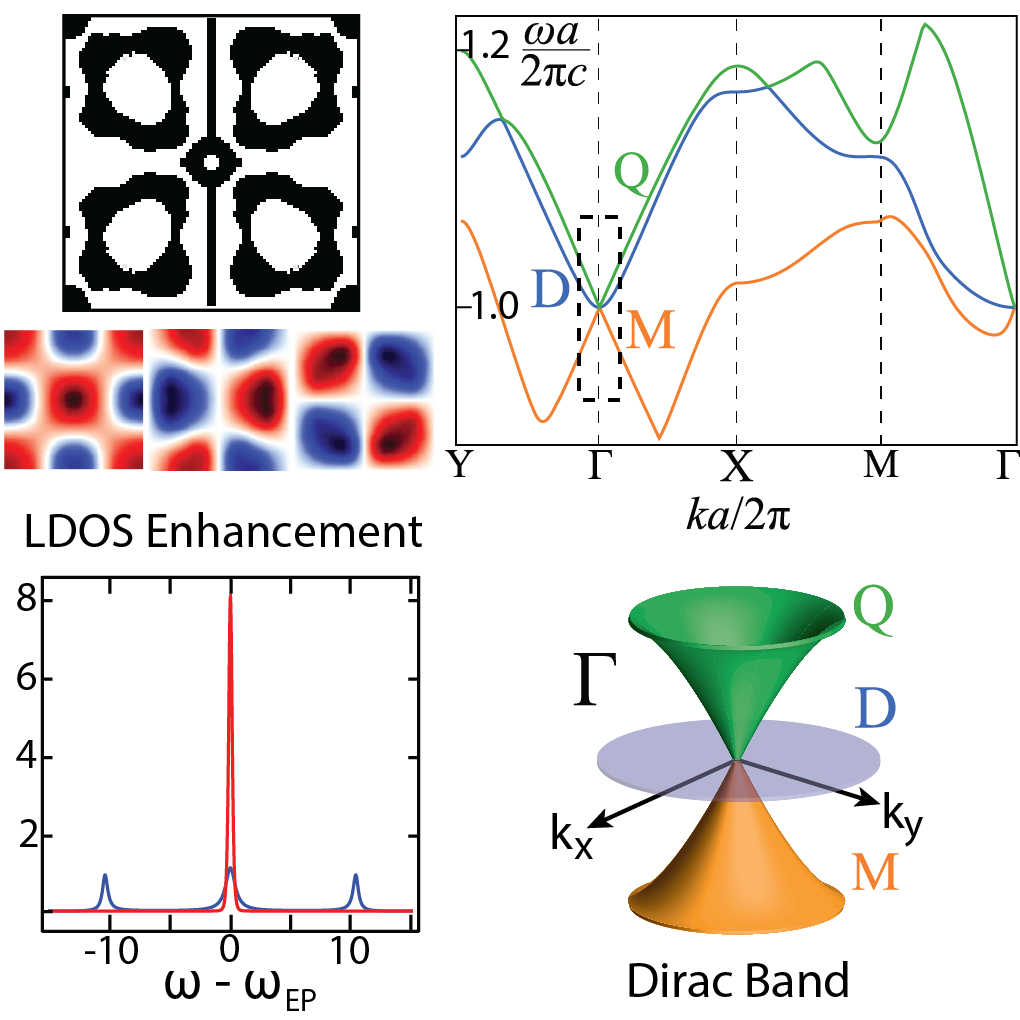}
  \vspace{0 pt}
  \caption{\textbf{Exceptional and topological photonics}: The figure
    depicts the band and modes of a 2d square lattice discovered by
    topology optimization\cite{lin2016enhanced}. At the $\Gamma$ point
    of the Brillouin zone, the monopole, dipole and quadruspole modes
    (labeled as M, D, and Q) coalesce creating an exceptional
    point. The resulting Dirac band structure and self orthogonality
    of the modes has been shown to strongly modify the qualitative
    characteristics of the local density of states, resulting in
    enhanced spontaneous emission and nonlinear effects.}
  \vspace{0 pt}
\end{figure} 

\section{Current and Emerging Applications} 
\label{sec:current_and_emerging_applications}
\noindent
\textbf{Nonlinear optics}: The utility of engineered resonances for
nonlinear phenomena is well documented\cite{ou1993enhanced}. Compared
to bulk media, resonators offer both longer interaction timescales and
higher field confinements, leading to increased nonlinear
interactions. Beginning with large-etalon cavities initially
considered in the mid nineteen sixties, and moving from millimeter- to
micron-scale whispering gallery mode resonators to the more recently
proposed wavelength-scale cavities\cite{lin2016cavity}, these ideas
have continued to be pushed to realize higher efficiencies (lower pump
powers), more compact architectures, and wider bandwidths (faster
operating timescales). At a finer level of detail, the physics of
nonlinear processes in wavelength-scale structures is well described
by a small set of parameters: the frequencies and decay rates (or
quality factors) of each resonance and the nonlinear overlap integrals
describing interactions between constituent electric fields
(generalizing the more commonly known, phase-matching condition
associated with propagating waves\cite{boyd2003nonlinear}).  These
properties fully characterize nonlinear interactions and must be
simultaneously tuned. Yet, while the statement of required conditions
is simple, the search for well suited structures remains both a
technical and conceptual challenge and none of the standard design
principles are readily applicable. While the creation of bandgaps
remains a valuable idea, they typically can only cover one of the
active frequencies. Further, even if this could be achieved, there is
no guarantee that it would result in desirable overlap
characteristics. Index-guiding structures can have high modal quality
factors and operate effectively over large
bandwidths\cite{furst2010naturally,bi2012high}, but require an unideal
tradeoff between mode confinement and radiative
losses. Plasmon-polariton resonances can provide excellent confinement
and field intensities but are saddled with the unavoidable presence of
material loss\cite{khurgin2015deal}, limiting the ultimate conversion
efficiency that can be achieved. Concurrently, for weak
nonlinearities, the distinct resonances of any structure must be
orthogonal, which weakens nonlinear overlap integrals and hence
interactions\cite{hashemi2009nonlinear}. To boot, devices based on
$\chi^{3}$ and higher order processes require amplitude-dependent
frequency corrections to account for cross- and self-phase modulations
that prove difficult to independently tune in few-parameter designs.\\ \\
The complexity implied in determining structures that simultaneously
achieve these various design objectives seems ideally suited to
inverse design techniques. As a conformation of this notion,
preliminary findings for $\chi^{2}$ second harmonic
generation\cite{lin2016cavity,sitawarin2017inverse}, and $\chi^{3}$
difference frequency generation\cite{lin2017topology} are presented
Fig.~2 (see figure for design descriptions). The three designs
depicted are found to have nonlinear figures of merit between one and
three orders better than any previously reported designs up to the
millimeter scale. Across the varied design paradigms considered
(layered micropillar cavities, multiring structures, metasurfaces,
etcetera), topology optimized structures are observed to have
systematically reduced structural symmetry, large quality factors, and
nonlinear overlaps. For the nonlinear processes considered, the
optimizations consistently find intuitive designs to be overly
simplistic in the sense that they do not make sufficient use of
interference to match the profiles of the interacting modes. From a
practical perspective, stronger overlaps are preferred to higher
quality factors since the former are less sensitive to fabrication
imperfections and offer greater speed (larger bandwidths).\\ \\
Finally, the realization of wavelength scale nonlinear devices for
$\chi^{2}$ and $\chi^{3}$ harmonic generation is a necessary step
towards the development of a variety of promising on-chip technologies
including low threshold lasers\cite{takahashi2013micrometre},
frequency combs\cite{okawachi2011octave},
imagining\cite{zipfel2003live}, supercontinuum
sources\cite{halir2012ultrabroadband},
spectroscopy\cite{moon1997absolute}, single photon
sources\cite{pelton2002efficient}, and quantum information
processing\cite{guo2016chip}. The early successes of topology
optimization in the nonlinear domain indicate that each of these
proposals may benefit substantially from incorporating inverse design
techniques.

\begin{figure*}[t!]
  \centering
  \includegraphics{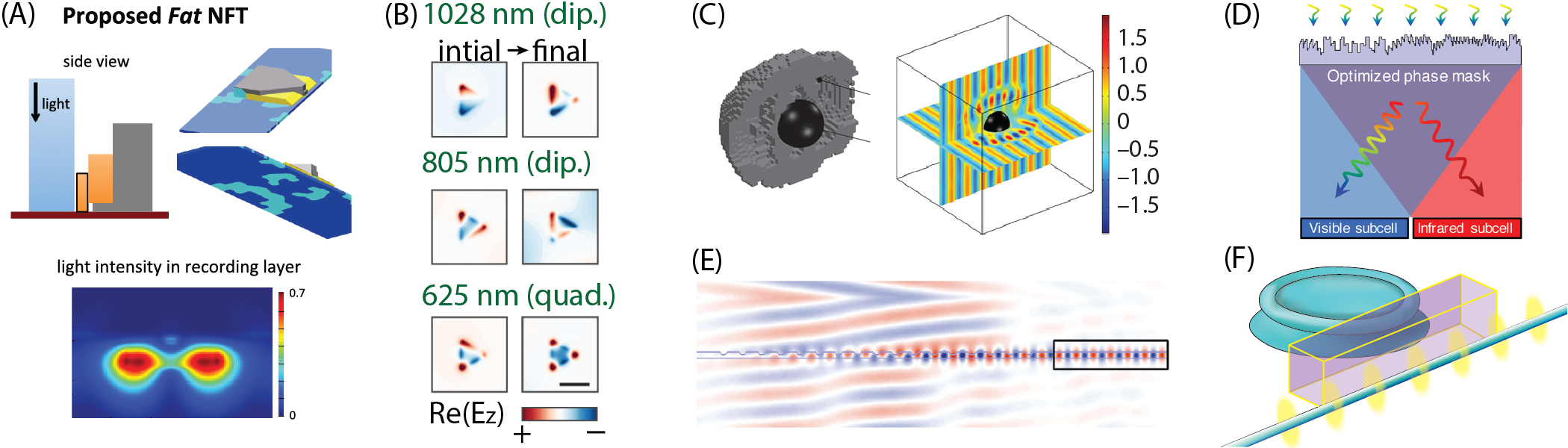}
  \vspace{0 pt}
  \caption{\textbf{Growth of applications}: The past three years have
    seen remarkable growth in variety of systems treated with
    computational adjoint methods. Panel (A) depicts a near-field
    transducer for heat-assisted magnetic recording, offering a 50\%
    reduction in self-heating compared with industry
    standards\cite{bhargava2015lowering}. (B) Illumination of an
    arbitrary nanoscale structure (triangle) with an optimally
    structured beam to increase optical torque, leading to a 20-fold
    enhancement\cite{lee2017non}. (C) An electromagnetic cloak,
    leading to order of magnitude reduction in total
    scattering\cite{deng2016topology}. (D) A schematic of an
    experimentally realized optimized structure for spectral splitting
    that achieves 69.5\% separation of the optical and infrared
    spectra, opening new directions for multi-bandgap
    photovoltaics\cite{xiao2016diffractive}. Panels (E) and (F)
    display conceived applications of inverse design structures to
    modal coupling: (E) free space coupling to a waveguide mode
    doubling the field amplitude compared to a traditional grating
    from Niederberger et al.\cite{niederberger2014sensitivity}; and
    (F) optimized coupling of power between a ring resonator and a
    waveguide.}
  \vspace{0 pt}
\end{figure*}
\noindent
\textbf{Exceptional and topological photonics}: Starting
from early investigations of 2d bandgaps, dispersion engineering has
consistently stood as of one of the strongest motivations for applying
inverse design in photonics. Spurred by the currently developing
understanding of topological properties in photonic systems, this
original inspiration has reemerged in the creation of exceptional
points. As in the case of nonlinear phenomena, manipulating the
nuanced role that structure plays in determining the exact
characteristics of theses features seems particularly aligned to
inverse approaches.\\ \\
Exceptional points occur in non-Hermitian problems (macroscopic
electromagnetics, acoustics, etc.) when two or more of the associated
complex eigenvalues coalesce, causing the basis to become
incomplete. The associated physical behavior is markedly different
from the more familiar accidental degeneracy encountered in Hermitian
systems. First, as modes approach an exceptional point, the remaining
eigenmode becomes self orthogonal\cite{heiss2012physics}, typically
quantified in terms of the diverging Petermann
factor\cite{berry2003mode}. Second, the existence of an exceptional
point also alters the the analytic properties of the Green's function,
introducing an additional pole with order equal to the degree of
coalescence\cite{pick2017general}.  The ramifications of these altered
response characteristics are linked to a long list of exotic optical
phenomena, including directional
transport\cite{regensburger2012parity}, anomalous
lasing\cite{peng2014loss}, and enhanced sensor
detection\cite{hodaei2017enhanced}. However, there are also
applications to more conventional photonic processes: exceptional
points have been predicted to enhance the efficiency of spontaneous
emission and frequency conversion\cite{pick2017enhanced}.\\ \\
Within the past decade, exceptional points have been designed using
three primary schemes: geometries involving gain and loss in coupled
resonators\cite{peng2014parity}, interacting
waveguides\cite{ruter2010observation} and currently, purely passive
photonic-crystal lattices\cite{zhen2016spawning}. Indirectly, this
variety indicates that the subset of systems where exceptional points
can occur is in fact quite large, and that with proper design tools
there may be enough freedom to engineer both degree and location. As a
promising inroad to the additional physics and design possibilities
offered by exceptional points, Fig. 3 describes the existence of a
coalescence of three eigenvalues occurring at the $\Gamma$ point of an
open $C_{2v}$ photonic crystal obtained by topology
optimization\cite{lin2016enhanced}. Two central results follow: First,
in showing that third-order exceptional points can be readily
engineered, the study strengthens the notion that exceptional points
are not bound to special geometries or material parameters. Second,
the authors show that exceptional points can be used to enhance the
local density of states at certain positions in the crystal. \\ \\
The results are also relevant in the burgeoning field of topological
photonics\cite{lu2014topological}. The Dirac bandstructure that
accompanies the creation of an exceptional point is a known precursor
to media with non trivial photonic topologies, encompassing
backscattering immune surface states\cite{wang2009observation},
topological insulators\cite{slobozhanyuk2017three}, and optical Weyl
points\cite{noh2017experimental}. 
Given the potential impact of realizing designer topological
properties in practical physical systems, the extension of inverse
design methods to deal with other key stepping stones such as chiral
modes, and omnidirectional Dirac cones, seems promising.

\begin{figure*}[t!]
  \centering
  \includegraphics{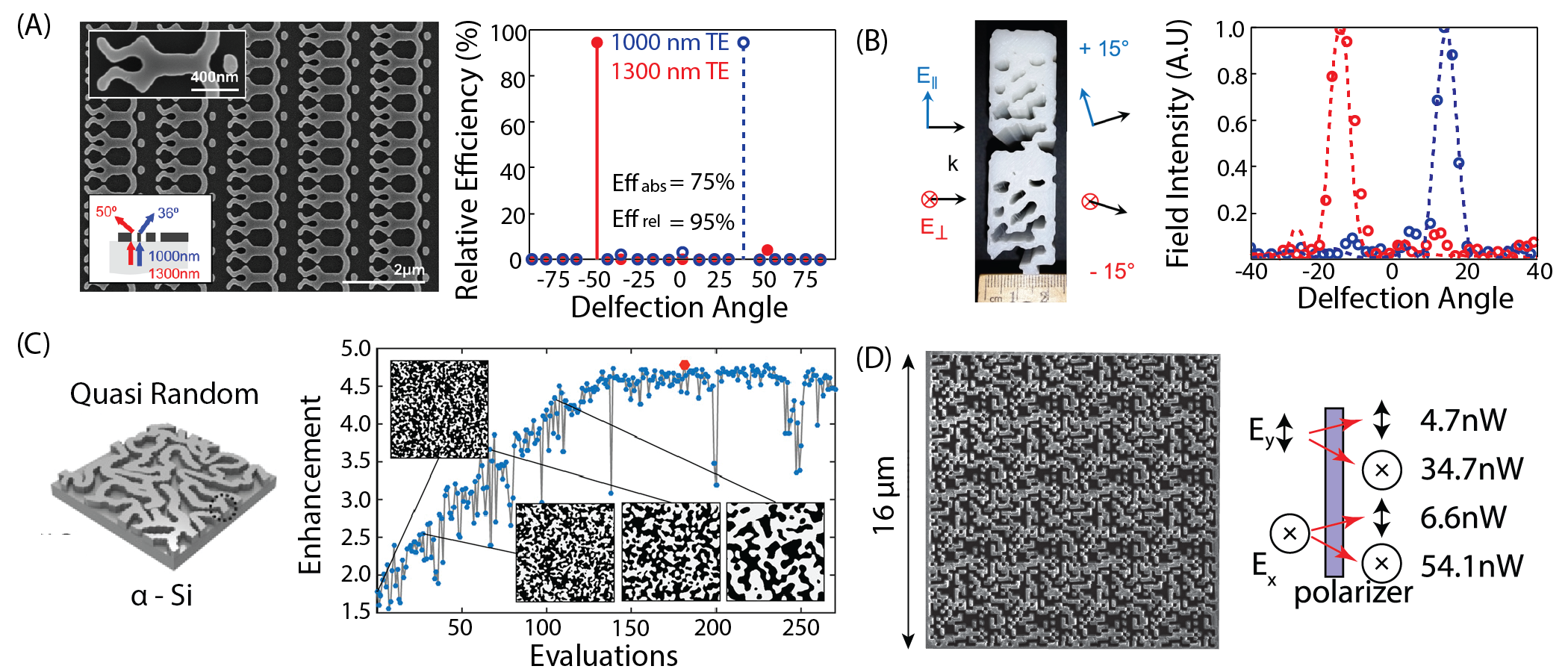}
  \vspace{0 pt}
  \caption{\textbf{Metasurface photonics}: The figure highlights four
    recent applications of inverse design to metaphotonics. (A) A
    metagrating capable of angularly separating 1000 nm and 1300 nm TE
    polarized light with 75\% absolute efficiency by Sell et
    al.\cite{sell2017large}. (B) A 3D polarization splitter designed
    for microwave applications ($\approx 26 - 33$ GHz) by Callewaert
    et al.\cite{callewaert2017inverse}. (C) An optimized quasi random
    amorphous silicon structure for enhancing absorption of the
    optical spectrum created by Lee et
    al.\cite{lee2017concurrent}. (D) A topology optimized polarizer
    with $\approx$ 90\% conversion efficiency conceived by Shen et
    al.\cite{shen2014ultra}.}
  \vspace{0 pt}
\end{figure*}
\noindent
\textbf{Nanoscale optics and metasurfaces}: Over the last several years,
large-scale optimization methods have begun to have a significant
impact on a diverse collection of problems in nano-optics and
metasurfaces. Representative selections are depicted in Figs. 4 and 5.\\ \\
Implementing a boundary inclusion optimization to determine the
characteristics of a slab waveguide, Bhargava and
Yablonovitch\cite{bhargava2015lowering} proposed a near field
transducer for heat assisted magnetic recording. The design has 50\%
less self heating than the current standard employed in
industry. Making use of the boundary element method, Lee et
al.\cite{lee2017non} investigated the optimization of electromagnetic
torques arising from incident optical fields on arbitrary
nanostructures. For the example triangular nanoparticle shown in
Fig. 4(B), the torque generated on the quadrupole mode was increased
by a factor of 20. As an example of their edge element method for
three dimensional volume optimizations, Deng and
Krovnik\cite{deng2016topology} applied topology optimization toward
the design of a single-material cloak for a perfect spherical
conductor, leading to an order of magnitude reduction in scattered
power. Each of these examples exemplifies a technologically relevant
area of photonics where complexity hampers direct application of
standard design principles.  Moreover, while in some cases there is
guidance on expected performance from existence of fundamental limits
(typically derived from physical constraints like energy conservation
or reciprocity)\cite{Yu10}, there are yet many situations (such as in
near-field or metasurface applications) where no such bounds exist or
are only beginning to emerge\cite{miller2015shape,arabi17}, and hence
where it is unclear what sort of performance can be achieved.\\ \\
Varied examples have also been reported for more traditional optical
problems such as diffraction, coupling, polarization control, and
absorption enhancement in constrained volumes, Fig. 4 (D)-(F) and
Fig.~5. In particular, a substantial number of promising results have
already been obtained in the context of metasurfaces.  The works of
Sell et al.\cite{sell2017large}, Callewaert et
al.\cite{callewaert2017inverse} and Shen et al.\cite{shen2014ultra}
have connected inverse design to the larger pursuit of flat optical
systems to replace the functionality of conventional optical
components\cite{yu2014flat}. Each of the three works presents a
general scheme and experimental realization for either highly
efficient diffraction, Fig. 5 (A)-(B), or polarization control, Fig. 5
(D), that can be applied to an assortment of particular problems. Much
as in the case of band structure, the findings of these studies open
broader questions about the breadth of phase and polarization control
that can occur per unit thickness in a structured medium (or a single
simply structured layer\cite{lin2017topology}). \\ \\
The metasurface concept also relates to the pressing demand to improve
solar energy capture. Two primary aspects which limit the efficiency
of traditional (simple) pn-junction designs are light trapping within
the volume where photovoltaic conversion
occurs\cite{yablonovitch1982intensity,garnett2010light} and the width
of the solar spectrum, which fundamentally limits the potential
conversion efficiency of any single bandgap. Any device design
offering improvement in either aspect is notable, especially if it
does not impose extreme fabrication difficulties and can be
implemented in silicon systems. While there have been many recent
works on inverse design for solar cell
applications\cite{ganapati2014light}, we highlight two particular
examples that address these two issues. The first issue was studied by
Shen et al.\cite{shen2014ultra} in 2014 with respect to the
quasi-random features that can be imposed on amorphous silicon surface
by wrinkle lithography. Conducting Fourier-based inverse design
(refereed to here as concurrent design) the authors were able to
enhance light-trapping by a reported factor of five over the spectral
range of 400 to 1200 nm. The second issue has been examined by Xiao et
al.\cite{xiao2016diffractive}, who in 2016 demonstrated a splitter
that physically separates optical and infrared wavelengths with 69.5\%
efficiency. By partitioning the solar spectrum in this way,
photovoltaics with different bandgaps can be placed in a side by side
configuration allowing for multiple bands of high-efficiency
operation. \\ \\
Simultaneously, significant progress has also been made on variations
of the question of mode couplers. For dense chip-scale integration
there is a clear need to limit the total optical footprint by handling
multi-frequency bands on a single waveguide. In opposition, there is
also a clear need to be able to access the information stored on these
different frequency bands independently. To meet both goals, devices
capable of high fidelity wavelength division multiplexing are
required. Adjoint optimized devices from Fresselen et
al.\cite{frellsen2016topology} and Piggot et
al.\cite{piggott2015inverse} for implementing this functionality in
areas of a few square microns at telecom wavelength, with sub 5 dB
transmission loss, are shown in the lower part of Fig.~1 and in
Fig.~6. Shen et al.\cite{shen2015integrated} and Mak et
al.\cite{mak2016binary} have come to similar findings. Also shown in
Fig. 6 is a three-port power splitter designed using a fabrication
tolerant algorithm and measured to have no worse than 23\%
transmission at any of its three output ports across its 1400 to 1700
nm operational bandwidth\cite{piggott2017fabrication}. Similar to
solar energy capture, any improvement in the components providing
these necessary functionalities potentially has far reaching
industrial impact. \\ \\
Finally, panels (E) and (F) in Fig.~4 present more speculative
applications; the free space coupling of light into a waveguide from
Niederberger et al.\cite{niederberger2014sensitivity} and the
optimized coupling of power between a ring resonator and a waveguide
at multiple frequencies using wavelength-scale elements. Both problems
are routinely dealt with in experimental settings. However, there are
surprisingly few high-efficiency techniques to couple light either
into nanophotonic structures from free space, or from a waveguide into
a cavity beyond adiabatic tapers.
\\ \\
\noindent
\textbf{Experimental challenges}: Since 2004, a variety of designs
have been experimentally demonstrated to illustrate the viability of
computational inverse methods. Ranging from bends and splitters for
photonic-crystal waveguides\cite{borel2004topology,
  piborel_el2005,jensen2011topology}, to passive components for
silicon photonic
circuits\cite{aypiggott_sr2014,piggott2015inverse,piggott2017fabrication,su2017inverse,frellsen2016topology}
and metasurfaces\cite{sell2017large, callewaert2017inverse},
operational devices exists in essentially every major domain of
photonics in which inverse design has been applied. Yet, to date, none
these structures have found broad industrial application.  The primary
cause of this incongruity is simple: nearly every device has been
fabricated using electron-beam lithography due to the small features
that occur naturally in current inverse algorithms. For industrial
applications, limiting fabrication time, and hence cost, requires
compatibility with photolithography; and while explicit constraints
imposing a minimal feature size can be implemented in one dimensional
designs with relative simplicity\cite{am_grat_arxiv2017,
  lsu_arxiv2017}, such approaches are considerably more difficult in
higher dimensions.

\begin{figure}
 \centering
  \includegraphics{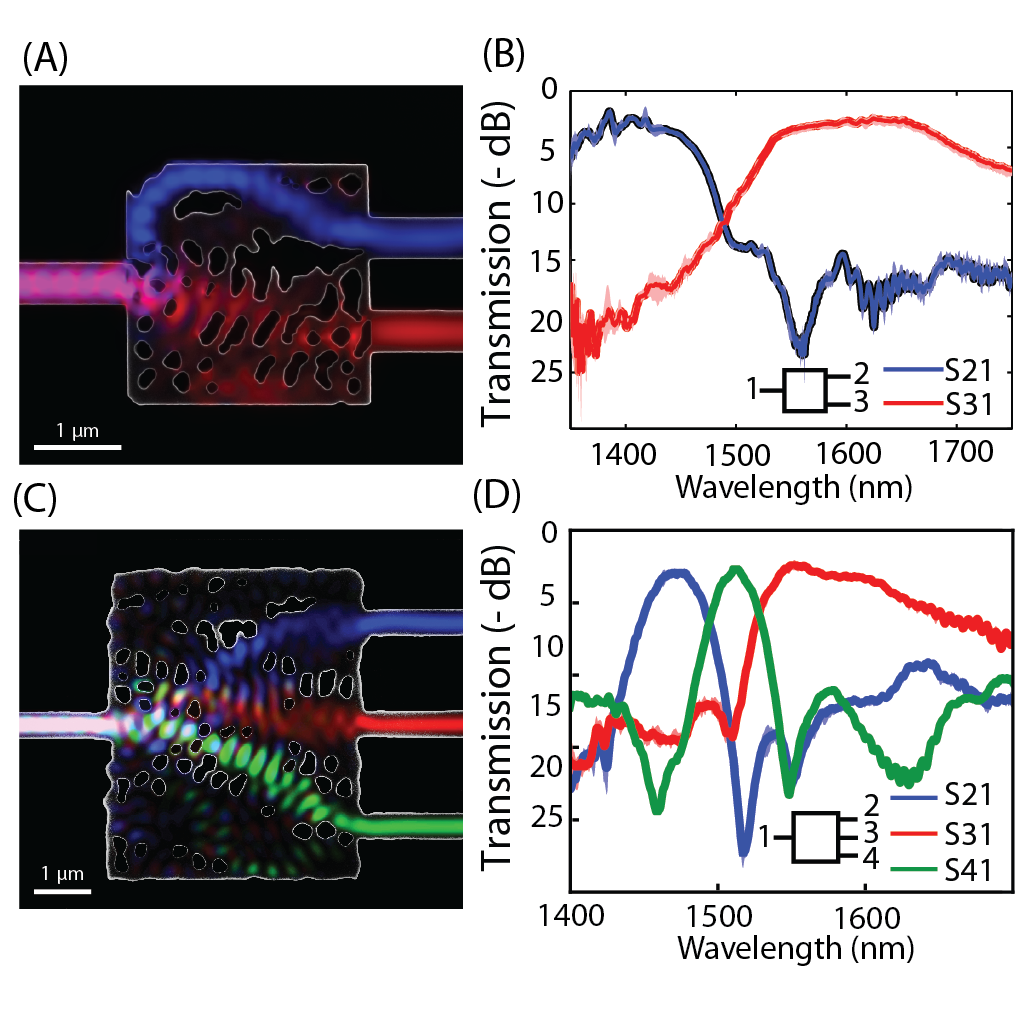}
  \vspace{-10 pt}
\caption{\textbf{Experimental inverse design}: The figure shows two
  SEM overlaid images with accompanying fields for narrowband (A)
  two- and (C) three-channel wavelength splitters, Piggot et
  al.\cite{piggott2015inverse,piggott2017fabrication}. The two-band
  splitter is designed to separate 1300 nm (blue) and 1550 nm
  (red). The three band splitter designed to separate
  $1500~\mathrm{nm}$ (blue), $1540~\mathrm{nm}$ (green), and
  $1580~\mathrm{nm}$ (red). Panels (B) and (D) show measured
  transmission spectra, validating the functionality of these
  devices.}
  \vspace{0 pt}
\end{figure}
\noindent
A conceptually simple solution to this challenge is to subdivide the
design region into pixels which are larger than the smallest
achievable feature size. After removing any intermediate gray
structures, the design is then assured to
fabricable\cite{,shen2015integrated,mak2016binary}. However, in trade,
this approach probes an overly limited design space, artificially
penalizing all smooth curves even if they do not require small
features. A more inclusive approach is to incorporate fabrication
constraints directly into the optimization problem. For topology
algorithms, this is accomplished by using convolutional filters to
smear out small features, and image dilation and erosion operations to
mimic fabrication
imperfections\cite{bslazarov_aam2016,frellsen2016topology,sell2017large}. For
boundary parameterized optimizations, simultaneously limiting the
minimum radius of curvature\cite{lalau-keraly_oe2013,
  am_bdy_arxiv2017}, and eliminating any gaps or bridges narrower than
a threshold width\cite{piggott2017fabrication,su2017inverse}, has been
shown to increase fabrication tolerance\footnote{Similar ideas can
  foreseeably be applied to make inverse designs robust to thermal
  variations\cite{jlu_oe2013}. Additionally, if materials with both
  positive and negative temperature coefficients of the refractive
  index are incorporated into the design, it should be possible to
  design almost completely athermal devices
  \cite{ssdjordjevic_oe2013}.}. These methods have been experimentally
tested for electron-beam lithography, and simulations indicated that
should also work reasonably well for photolithography when using
optical proximity correction\cite{wwang_spie2017}. Yet as promising as
these results are, they are not robust to process variations in
photolithography, such as defocusing and dosage errors, and finding
methods to cope with these additional complications remains an open
problem.\\ \\
Finally, the physical size of practical aperiodic devices that can be
currently treated with inverse design methods is limited by the
computational cost of the fully-vectorial 3d simulations needed to
accurately model their performance. Dozens to hundreds of simulations
are required to design a single device, which becomes prohibitively
expensive as design domains expand. This limits the type of questions
that can be meaningfully treated, and makes it difficult to inverse
design interfaces with large structures such as single-mode optical
fibers. In this light, further improvements in computational
approaches (such as iterative solvers) that underpin current inverse
methods have the potential to vastly expand complementary application
boundaries.

\section{Summary Outlook} 
\label{sec:summary_outlook}
\noindent
Applications have always served as the vital spark for progress in
inverse design, and from this fact alone the outlook for the
application of these principles in nanophotonics is positive. There is
both a clear set of mature, clearly formulated problems in areas such
as chip-scale integration and cavity design that remain open; as well
as a range of new areas of application such as energy capture and
nonlinear device design where only promising preliminary work has been
done. Beyond the areas we have outlined in the previous sections,
inverse design principles appear to offer a new perspective for
understanding fluctuation physics and near-field optics. Although
currently limited to one dimension\cite{jin2017overcoming}, the
application of topology optimization to find optimally efficient heat
transfer systems\cite{bhargava2015lowering} promises to have practical
and theoretical impact in building further understanding of the
practical limits of heat transfer\cite{miller2015shape}. Extending
inverse design to active devices such as modulators and lasers, which
are often the performance limiting components of optical systems,
would also be extremely useful.\\ \\
A number of key improvements would enable the widespread usage of
inverse design methods in practical applications. First and foremost
is improving the robustness of designs to handle process variations in
photolithography, which would enable high throughput
fabrication. Parallel to this computational focused tract, advancement
in nanoscale lithography\cite{kasahara2016recent} appear poised to
enlarge the landscape of fabricable structures to include a larger
subset of the intricate multiscale features and permittivity
gradients\cite{urness2015arbitrary} that ubiquitously appear in
inverse algorithms. Inverse design methods can, at least in principle,
explore the full space of fabricable devices. It thus becomes a very
meaningful question to ask: what is the maximum theoretical
performance of an optical device? More specifically, for a given
design area, minimum feature size, and selection of materials, what is
the ultimate achievable performance of an optical device for a
particular function?  Establishing such theoretical bounds on the
performance of optical devices would help guide future work in all of
photonics. \\ \\
Finally, improvements to the underlying simulations and optimization
algorithms could enable design of larger devices, greatly improving
the breadth and scope of problems that can be tackled by inverse
design. Along these lines, several recent works have begun exploring
applications of machine learning in
nanophotonics\cite{shen17,zibar17,turduev17}, paving the way for
instance to developments in the area of fast, iterative Maxwell
solvers. In the enduring quest for optimal photonic designs, the
widespread integration of inverse design tools seems not only sensible
but unavoidable.

\bibliography{revPID}

\end{document}